# Novel synthesis method of nonstoichiometric $Na_{2-x}IrO_3$. Crystal structure, transport and magnetic properties.

*Katharina Rolfs\*[†], Ekaterina Pomjakushina[†], Denis Sheptyakov[‡], and Kazimierz Conder[†]*

[†] Laboratory for Scientific Development and Novel Materials, Paul Scherrer Institute, Villigen PSI, 5232 Switzerland

[‡] Laboratory for Neutron Scattering and Imaging, Paul Scherrer Institute, Villigen PSI, 5232 Switzerland

**Abstract**

Transition metal oxides with 4d or 5d metals are of great interest due to the competing interactions, of the Coulomb repulsion and the itineracy of the d-electrons, opening a possibility of building new quantum ground states. Particularly the 5d metal oxides containing Iridium have received significant attention within the last years, due to their unexpected physical properties, caused by a strong spin orbit coupling observed in Ir(IV). A prominent example is the Mott-insulator $Sr_2IrO_4$. Another member of this family, the honeycomb lattice compound $Na_2IrO_3$, also being a Mott-insulator having, most probably, a Kitaev spin liquid ground state. By deintercalating sodium from $Na_2IrO_3$, we were able to synthesize a new honeycomb lattice compound with more than 50% reduced sodium content. The reduction of the sodium content in this layered compound leads to a change of the oxidation state of iridium from +IV to +V/+VI and a symmetry change from *C2/c* to *P-3*. This goes along with significant changes of the physical properties. Besides the vanishing magnetic ordering at 15K, also the transport properties changes and instead insulating semiconducting properties are observed.

**Introduction**

In the field of strongly correlated electron systems significant attention has been drawn to 4d and 5d based transition metal oxides (TMO). The spin orbit coupling (SOC) observed in these systems becomes non-negligible and leads to new exotic ground states, such as the Mott insulating state in $Sr_2IrO_4$ [1]. The interest is focused, among others, on Mott insulators with properties approaching a metal- insulator transition. It was suggested that a spin liquid ground state can be find in such compounds, which farther can lead to new phases, with interesting electronic properties such as high-$T_c$ superconductivity in cuprates[2]. Several iridates, like $Na_2IrO_3$ and $Li_2IrO_3$, have been discovered to be quantum frustrated Mott insulators [3]. In these compounds, the SOC is the compelling property, which stabilizes the insulating state even though the onsite Coulomb interactions are relatively weak. Another interesting attribute of $Na_2IrO_3$ is its honeycomb structure [4,5], enabling the realization of the exactly solvable spin model with spin liquid ground state proposed by Kitaev [6]. It was suggested that these materials contain the necessary anisotropic exchange interactions ($J_K$) and that the spin liquid ground state resists the small but always present isotropic exchange (J) [7]. However $Na_2IrO_3$ and $Li_2IrO_3$ order magnetically at 15 K, even though this transition temperature is low compared to the characteristic exchange energy [8]. To access the spin liquid phase, the reduction of these isotropic Heisenberg terms or the increase of the anisotropic interactions is necessary. Reuther et al. [7] suggested that the reduction of the chemical pressure along the c-axis can increase $J_K$ and therefore induces the spin glass behavior. This can be achieved either by exerting pressure in the *ab*-plane or substituting Na by smaller Li-ions. Another possible chemical modification of $Na_2IrO_3$ is a hole doping. This was predicted to lead to a topological superconductivity and a Fermi liquid behavior in the vicinity of the Kitaev spin liquid phase [8,9]. However, the only way to introduce holes into $Na_2IrO_3$ is introducing sodium deficiency. The direct synthesis of $Na_{2-x}IrO_3$, would result in a mixture of $IrO_2$ and $Na_2IrO_3$. As shown by Takeda et al. [10] for $Na_{0.7}CoO_2$, it is possible to oxidize the material by deintercalating the sodium positioned in between the $CoO_2$ layers. Applying the same chemical oxidation process, it was also possible to oxidize the layered compound $Na_2IrO_3$ to $Na_{0.8}IrO_3$. Here we will show a crystallographic study of this new compound, accompanied by magnetization and transport measurements, showing the significant influence of the change of sodium stoichiometry on physical properties.

**Experimental**

Polycrystalline $Na_2IrO_3$ was synthesized by a solid state reaction using $Na_2CO_3$ and $IrO_2$. A stoichiometric mixture of both powders was heated at 900 °C for 48 h. To ensure a clean compound, the product has been measured by X-ray diffraction (XRD) at room temperature using a Bruker D8 diffractometer in Bragg Brentano Geometry with Cu Kα radiation.

The $Na_2IrO_3$ powder was added to a $Br_2/CH_3CN$-solution for 6 h, 24 h and 48 h and subsequently rinsed with water. The dry product was studied by TG/DTA Thermal Analysis equipped with mass spectrometer. With these, thermal stability of the compound could be checked and possible phase transitions above room temperature detected. The crystallographic structure was determined by the neutron powder diffraction at 50 K and 2 K with wavelength λ = 1.49 Å at the High Resolution Powder Diffractometer HRPT[11] at SINQ, PSI in Switzerland. The magnetic properties were measured at a Quantum Design Magnetic Property Measurement System (MPMS), the resistivity and heat capacity measurements were done using the Physical Property Measurement System (PPMS).

**Result and Discussion**

**Structural change and thermal stability, crystal water**

Room temperature XRD measurements of $Na_2IrO_3$ samples, chemically oxidized for 6 h, 24 h and 48 h are showing very similar pattern which is clearly different from those observed for the parent phase $Na_2IrO_3$. The diffraction patterns could be indexed in a trigonal unit cell by DICVOL04 with the lattice constants for the sample deintercalated for 24 h, $a$ = 5.230 Å and $c$ = 4.848 Å. The lattice constants for the samples deintercalated for 6 h and 48 h are very similar, as it can be seen in Table 1. Using the Rietveld refinement method, the space group of all three specimens was determined as *P-3* instead of the monoclinic *C2/c* symmetry of the parental phase $Na_2IrO_3$[5]. Furthermore, the refinement revealed a significant change in the Na-concentration in the samples from $Na_2IrO_3$ to approximately $Na_{0.8}IrO_3$. However, the amount of sodium removed from the compounds appears to be independent of the time interval of deintercalation chosen here, since all three specimens shown very similar stoichiometry (see Table 1). The sample, deintercalated for 24h was also measured by neutron diffraction at 2 K and 50 K (see Fig. 1), showing no structural phase transition on cooling down to 2 K. Structure refinement could be made with a very high quality as obtained $R_f$ – factors, $R_F = 100 \frac{\sum_h |'F_{obs,k}'-F_{calc,h}|}{\sum_h |'F_{obs,h}'|}$, of 4.81 for the neutron data at 2 K, and 3.72 at 50 K are showing. The refinement confirms the significant

reduction of the Na-content in the sample from $Na_2IrO_3$ to $Na_{0.8}IrO_3$ (see Table 2). This change in stoichiometry goes along with the oxidation of Ir from +IV to a mixed +V/+VI state.

Further the refinement of the neutron data shows, that the thermal parameters of sodium exhibit a high anisotropy at 2 K and 50 K along the $c$ – axis. The discrepancy still observed between the calculated and observed intensities is present due to a lowered crystallinity of the specimens. Since the samples have been rinsed with water, it can be expected to have also guest water molecules within the sodium layers (see Scheme 1). This can explain the measured thermal anisotropy of the sodium as well as the reduced crystallinity of the samples.

Thermogravimetic measurements made on heating (1 K/min) in Helium (see Fig. 2) are showing a decrease of the mass of 4.68% within temperature range 700-825K. At these temperatures a release of both water and oxygen from the sample was detected by the mass spectrometer. The release of water at such high temperatures elucidates, that the water is chemically bounded in the structure (crystal water), as discussed above. Additionally, the observed oxygen release indicates decomposition of the compound to $IrO_2$ and $Na_2O$, which was confirmed by the XRD measurements. Based on the decomposition products, the reaction equation can be written as

$$Na_{0.8}IrO_3 \cdot (H_2O)_x \rightarrow 0.4\ Na_2O + IrO_2 + 0.3\ O_2 \uparrow + x\ H_2O \uparrow$$

From the mass change we can roughly estimate the amount of crystal water, to be 0.97 % of the mass, which is equal to 0.14 mol. In a second reaction step starting at 1000K, the $IrO_2$ is reduced to Ir and $O_2$. The attempt to increase the crystallinity of the samples, by annealing below the decomposition temperature at 723 K for 24 h resulted in the decomposition as well.

**Magnetic properties**

Magnetic susceptibility measurements of all deintercalated samples showed similar magnetic properties. Instead of an antiferromagnetic transition present in the parent phase at 15 K, all specimens show a paramagnetic behavior down to 2 K. For the sake of clarity only the data for the 24h-deintecalated sample are plotted in Figure 3. The linear fit of the inverse susceptibility between 250 K and 300 K gives an effective magnetic moment $\mu_{eff} = 1.84\mu_B$ which is consistent with a mixture of iridium on fifth and sixth oxidation states, with the expected effective moments of 2.82 $\mu_B$ for Ir(V) and 1.73 $\mu_B$ for Ir(VI). However the effective moment found in all three samples is smaller than expected for the overall stoichiometry found by structural studies. This can be caused by a small intrinsic itinerant moment, but considering the low temperature Curie tail in all measured samples, the reduced moment is probably caused by small proportion of paramagnetic impurities. With a Curie Weiss Temperature $\theta$ of -1347 K, the system seems to be highly frustrated. However the high Curie temperature could be partially caused by the change of the moment with temperature, since the material shows semiconducting behavior.

As it is listed in table 3, the samples deintercalated for 6 h and 48 h respectively, show a similar magnetic moment and $\theta$, supporting the result, that within the time frame of 6h to 48 h no further significant amount of sodium is removed. Since the Curie temperature is very high in all the samples, and the susceptibility was measured up to 300 K, the effective magnetic moment cannot be properly fit, which can explain the small difference within the samples, assuming the samples to have very similar stoichiometry.

**Transport measurements**

Consistently with the absence of magnetic ordering or structural phase transition at low temperature, the heat capacity measurements show no peak in the measured temperature range from 2 K up to 300 K. As stated in the Introduction, the strong spin orbit coupling of Iridium in the oxidation state +IV can lead to the Mott insulator state in several iridates. This state is observed in the parental phase $Na_2IrO_3$. As discussed before, the deintercalation leads to the oxidation of the Ir(+IV) to a mixed Ir(+V/+VI) state, changing also the ground state of $Na_{0.8}IrO_3$. Bremholm et al., found the high pressure $NaIrO_3$ containing pentavalent Iridium, has a nonmetallic behavior with a variable range hopping.[12] The resistivity measurements of $Na_{0.8}IrO_3$ showed a much lower resistivity of 0.0028 Ω cm at room temperature and 0.0427 Ω cm at 2 K than found in $NaIrO_3$ (400 Ω cm at 2 K). Furthermore a semiconductor behavior was revealed (see Figure 4) and could be fitted in the high temperature region between 260 K and 310 K by the Arrhenius law $\sigma(T) = \sigma_0 e^{(-E_a/kT)^m}$ with m=1, with a band gap of 0.083 eV.

**Conclusion**

In conclusion we could present a new method to synthesize a compound with Iridum at higher oxidations state than +IV at ambient pressure. The deintercalation of sodium from $Na_2IrO_3$ results in a significant reduction of the sodium content enabling a new way of oxidizing Ir(IV) Ir(V) and Ir(VI). This goes along with the formation of a new compound with a trigonal crystal structure (space group P-3). Independent of the time within the chosen timeframe of 6 h to 48 h, the deintercalation process reached its "equilibrium" resulting in the same final stoichiometry of $Na_{0.8}IrO_3$. We could show that the lattice incorporates small amount of water molecules during the rinsing process. With the deintercalation and the resulting oxidization of the Iridium from +IV to a mixed +V/+VI state, the samples show no magnetic ordering down to 2 K. Furthermore the ground state changes from a Mott insulating state to a semiconductor behavior with a band gap of 0.083 eV.

**Corresponding Author**

*Katharina Rolfs

**Author Contributions**

The manuscript was written through contributions of all authors. All authors have given approval to the final version of the manuscript.

**Acknowledgment**

This work is based on experiments performed at the Swiss spallation neutron source SINQ, Paul Scherrer Institute, Villigen, Switzerland. We acknowledge the allocation of the beam time at the HRPT diffractometer of the Laboratory for Neutron Scattering and Imaging (PSI, Switzerland). The authors thank SNF Sinergia project "Mott physics beyond Heisenberg model" for the support of this study.

**Abbreviations**

XRD X-ray diffraction; SOC Spin orbit Coupling; TMO Transition metal oxide

**References**

(1) Kim, B.; Jin, H.; Moon, S.; Kim, J.-Y.; Park, B.-G.; Leem, C.; Yu, J.; Noh, T.; Kim, C.; Oh, S.-J.; Park, J.-H.; Durairaj, V.; Cao, G.; Rotenberg, E. *Phys. Rev. Lett.* **2008**, *101* (7), 1–4.

(2) Kim, J.; Casa, D.; Upton, M. H.; Gog, T.; Kim, Y.-J.; Mitchell, J. F.; van Veenendaal, M.; Daghofer, M.; van den Brink, J.; Khaliullin, G.; Kim, B. J. *Phys. Rev. Lett.* **2012**, *108* (17), 177003.

(3) Singh, Y.; Gegenwart, P. *Phys. Rev. B* **2010**, *82* (6), 064412.

(4) Chaloupka, J.; Jackeli, G.; Khaliullin, G. *Phys. Rev. Lett.* **2010**, *105* (2), 027204.

(5) Choi, S. K.; Coldea, R.; Kolmogorov, A. N.; Lancaster, T.; Mazin, I. I.; Blundell, S. J.; Radaelli, P. G.; Singh, Y.; Gegenwart, P.; Choi, K. R.; Cheong, S.-W.; Baker, P. J.; Stock, C.; Taylor, J. *Phys. Rev. Lett.* **2012**, *108* (12), 127204.

(6) Kitaev, A. **2005**.

(7) Reuther, J.; Thomale, R.; Trebst, S. *Phys. Rev. B* **2011**, *84* (10), 100406.

(8) Mei, J.-W. *Phys. Rev. Lett.* **2012**, *108* (22), 227207.

(9) You, Y.-Z.; Kimchi, I.; Vishwanath, A. *Phys. Rev. B* **2012**, *86* (8), 085145.

(10) Takada, K.; Sakurai, H.; Takayama-Muromachi, E.; Izumi, F.; Dilanian, R. A.; Sasaki, T. *Nature* **2003**, *422* (March), 53–55.

(11) Fischer, P.; Frey, G.; Koch, M.; Ko, M.; Pomjakushin, V.; Schefer, J.; Thut, R.; Schlumpf, N.; Bu, R.; Greuter, U.; Bondt, S.; Berruyer, E. *Phys. B* **2000**, *276*, 146–147.

(12) Bremholm, M.; Dutton, S. E.; Stephens, P. W.; Cava, R. J. *J. Solid State Chem.* **2011**, *184* (3), 601–607.

(13) Toth, S.; Lake, B. *J. Phys. Condens. Matter* **2015**, *27*, 166002.

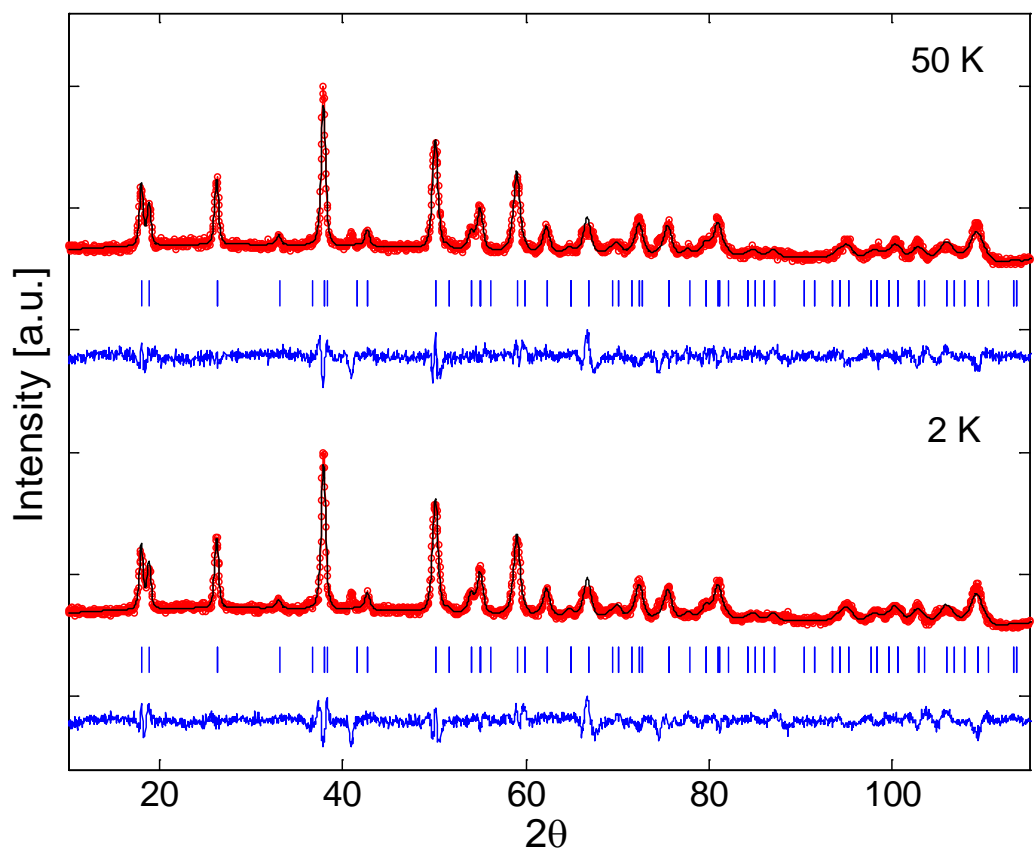

**Figure 1:** Rietveld Refinement with FullProf of the neutron powder diffractogram of $Na_{0.8}IrO_3$ measured at HRPT, SINQ at 2 K and 50 K.

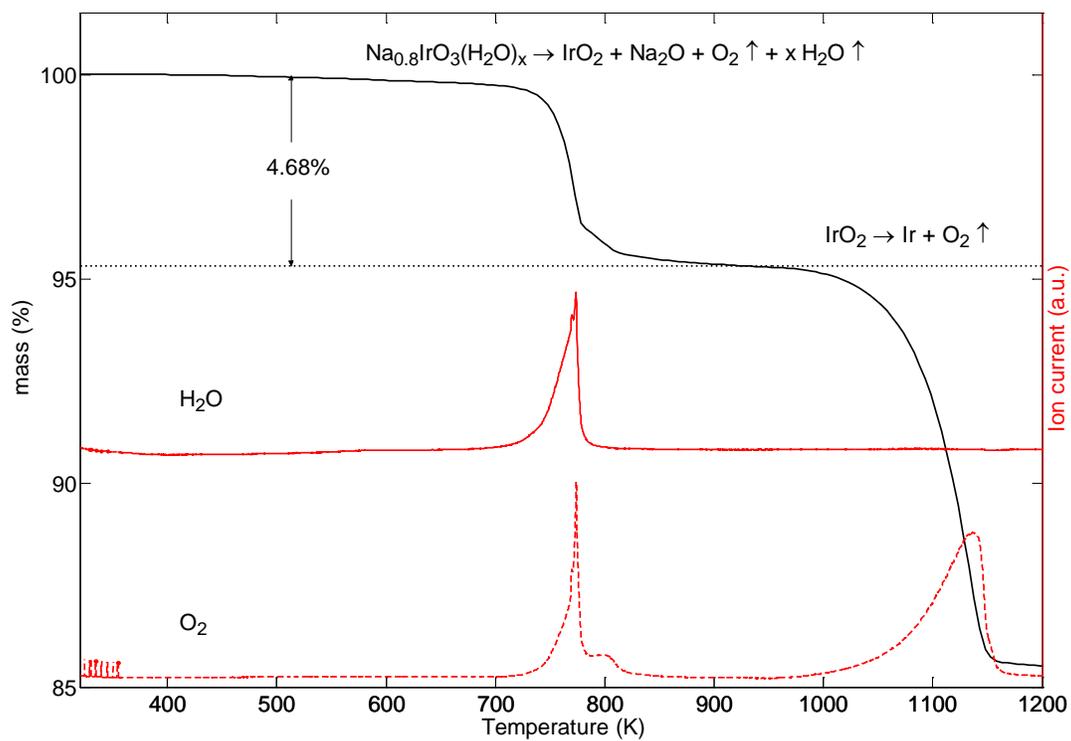

Figure 2: Thermogravimetric measurement of $Na_{0.8}IrO_3$ deintercalated for 24h. The black curve shows the mass change versus temperature. The red solid line shows the mass spectrometry signal corresponding to $H_2O$, the red dotted line corresponding to $O_2$.

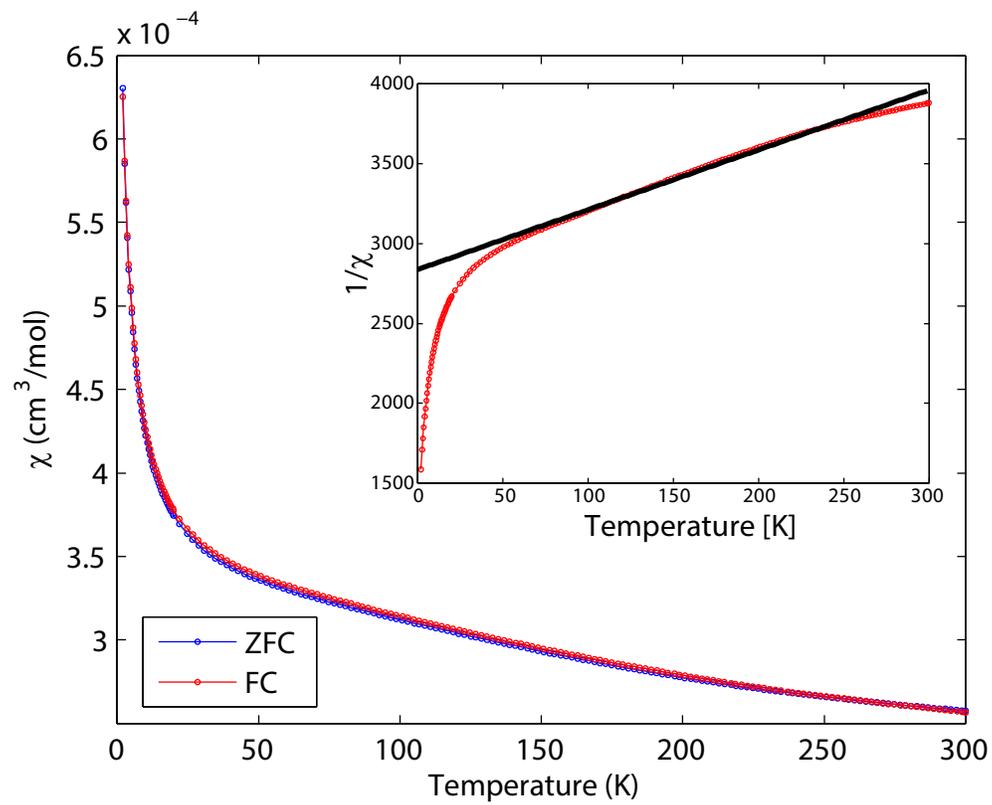

Figure 3: Susceptibility data of $Na_{0.8}IrO_3$, deintercalated for 24 h, measured at 1T. The inset shows the fitted inverse susceptibility.

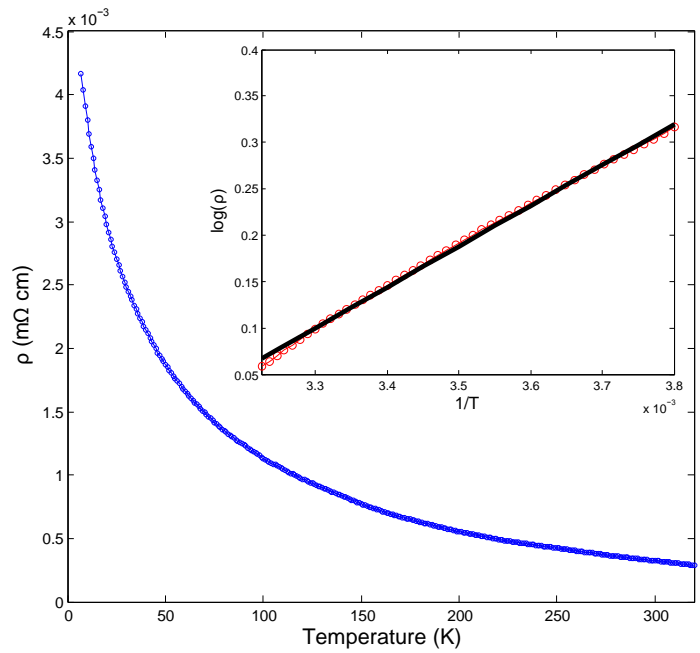

Figure 4: Resistivity of $Na_{0.8}IrO_3$ deintercalated for 24 h, indicating a non-metallic behavior. The inset shows the linear fit of log(p) vs 1/T with the Arrhenius law $\sigma(T) = \sigma_0 e^{(-E_a/kT)^m}$ with m=1.

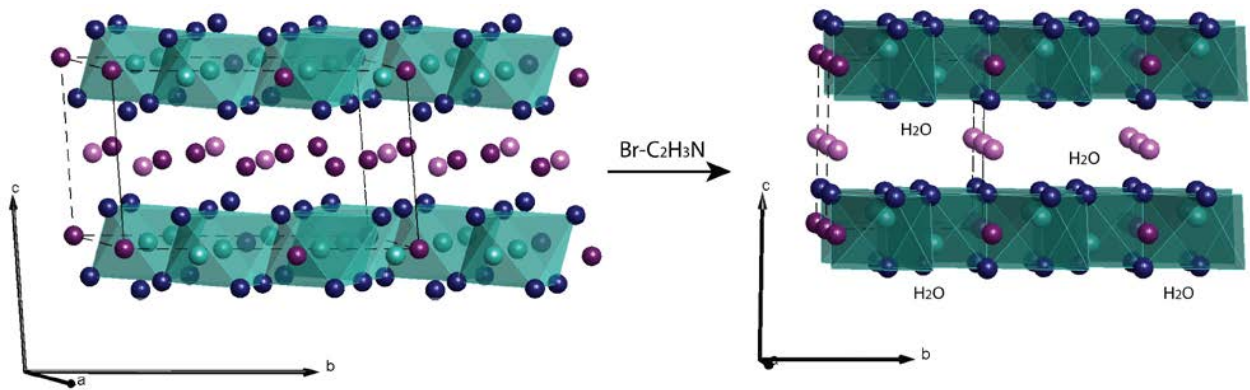

**Scheme 1.** Scheme of the parental compound $Na_2IrO_3$ (left structure) being converted to $Na_{0.8}IrO_3$ (right structure) by deintercalation of sodium plotted with SpinW[13]. The dark blue spheres illustrate the Oxygen atoms, the turquoise one, the Iridium atoms in the Oxygen tetraeders. The pink and purple spheres show the position of the two Na atoms. In $Na_{0.8}IrO_3$ these positions are also occupied by water molecules and the sodium shows a high anisotropy along the c-axis.

Table 1. Stoichiometry, cell parameters and $R_f$ factor of Rietveld refinement of $Na_{2-x}IrO_3$ deintercalated for 6 h, 24 h and 48 h.

| Time [h] | Stoichiometry | a [Å] | c [Å] | $R_f$ |
|---|---|---|---|---|
| 6 | $Na_{0.81}IrO_3$ | 5.236(5) | 4.858(3) | 2.81 |
| 24 | $Na_{0.76}IrO_3$ | 5.230(4) | 4.848(2) | 2.26 |
| 48 | $Na_{0.84}IrO_3$ | 5.232(2) | 4.848(6) | 2.52 |

Table2 . Unit-Cell, Positional and Displacement Parameters for $Na_{0.8}IrO_3$ oxidized for 24h with the trigonal P-3 Space Group at 2 K and 50 K, obtained from Rietveld Refinement of Neutron powder diffraction data and at room temperature from X-ray diffraction data

|  | T = 2 K | T = 50 K | T = 300K |
|---|---|---|---|
| $a$ (Å) | 5.248(0) | 5.247(8) | 5.230(4) |
| $c$ (Å) | 4.754(6) | 4.755(2) | 4.848(2) |
| $V$ (Å$^3$) | 113.405 | 113.412 | 114.865 |
| **Na1** (0,0,0) | | | |
| B (Å$^2$) | 0.81(6),1.4(2) anisotropic | 1.49(6), 0.9(3) anisotropic | -0.27(4), -0.0(5) anisotropic |
| Occ | 0.5(5) | 0.6(4) | 0.6(3) |
| **Na2** (0,0,1/2) | | | |
| B (Å$^2$) | -0.002(7), -0.03(7) anisotropic | -0.00(6), 0.02(4) anisotropic | 0.01(5), -0.02(6) anisotropic |
| Occ | 0.96(1) | 0.96(4) | 0.89(8) |
| **Ir** (1/3, 2/3, z) | | | |
| z | -0.014 | -0.01(1) | 0.00(03) |
| B (Å$^2$) | 0.01 | 0.2(8) | 0.36(7) |
| **O** (x,y,z) | | | |
| x | 0.62(7) | 0.62(6) | 0.67(1) |
| y | 0.62(5) | 0.62(5) | 0.64(9) |
| z | 0.216(9) | 0.216(7) | 0.17(9) |
| B (Å$^2$) | 0.98 | 0.83(8) | 0.30(5) |
| **R$_f$** | 4.81 | 3.72 | 2.26 |

Table 3: The magnetic moment $\mu_{eff}$ and Curie Weiss temperature $\theta$ of $Na_2IrO_3$ deintercalated for 6 h, 24 h and 48 h.

| $Na_2IrO_3$ deintercalated for | $\mu_{eff}$ [$\mu_B$] | $\theta$ [K] | C [$cm^3 K/mol/Oe$] |
|---|---|---|---|
| 6 h | 1.95 | -1559 | 0.4785 |
| 24 h | 1.84 | -1347 | 0.4244 |
| 48 h | 1.68 | -1186 | 0.3545 |